\documentclass[twocolumn,showpacs,preprintnumbers,amsmath,amssymb]{revtex4}

\usepackage{graphicx}
\usepackage{dcolumn}
\usepackage{bm}
\usepackage[usenames]{color}
\usepackage[normalem]{ulem}

\begin{document}

\preprint{APS/123-QED}
\title{Two-fold symmetry flattens Dirac cone of surface state at W(110) }%

\author{K. Miyamoto$^{1}$}
\email{kmiyamoto@hiroshima-u.ac.jp}
\author{A. Kimura$^{2}$}

\author{T. Okuda$^{1}$}
\author{K. Shimada$^{1}$}
\author{H. Iwasawa$^{1}$}
\author{H. Hayashi$^{1}$}
\author{H. Namatame$^{1}$}
\author{M. Taniguchi$^{1,2}$}
\author{M. Donath$^{3}$}
\affiliation{
$^{1}$Hiroshima Synchrotron Radiation Center, Hiroshima University, 2-313 Kagamiyama, Higashi-Hiroshima 739-0046, Japan
 }

\affiliation{
$^{2}$ Graduate School of Science, Hiroshima University, 1-3-1 Kagamiyama, Higashi-Hiroshima 739-8526, Japan
}

\affiliation{
$^{3}$ Physikalisches Institut, Westf\"alische Wilhelms-Universit\"at M\"unster, Wilhelm-Klemm-Strasse 10, 48149 M\"unster, Germany
}

\date{\today}

\begin{abstract}
 The $C_{2v}$ symmetry of the W(110) surface influences strongly the spin-polarized Dirac-cone-like surface state within  a spin-orbit-induced symmetry gap.
We present a detailed angle-resolved photoemission study with $s$- and $p$-polarized light along three different symmetry lines. 
The Dirac-cone-like feature appears along $\overline{\Gamma \rm H}$  and $\overline{\Gamma \rm S}$, while it is strongly deformed along $\overline{\Gamma \rm N}$. 
A two-fold $\Sigma_{3}$ symmetry of the $d$-type surface state is identified from photoemission experiments using linear polarized light.
Our results are well described by model calculations based on an effective Hamiltonian with $C_{2v}$ symmetry including Rashba parameters up to third order.  
The flattened Dirac cone of the surface state is caused by hybridization with bulk continuum states of $\Sigma_{1}$ and $\Sigma_{2}$ symmetry. 
The spin texture of this state obtained from the model calculations shows a quasi-one dimensional behavior.
This finding opens a new avenue in the study of $d$-electron-based persistent spin helix systems and/or weak topological insulators.

\end{abstract}

\maketitle
 Topological insulators and Rashba systems have attracted great attention with regard to dissipationless spin current transport without external magnetic fields~\cite{Bychkov84,Datt90, Nita97, Miron10}.
These materials possess spin-split energy-band structures induced by a combination of strong spin-orbit interaction and broken spatial inversion symmetry.
They have been extensively studied by angle-resolved photoemission spectroscopy (ARPES) with spin resolution and first-principles calculations in recent years \cite{Hasan10, Simon10, Shitade09, Hoesch04, Hirahara08, Kadono08, Dil09, Fu09, Souma11, Kuroda10, Hsieh09, Kuroda101,Hochstrasser02, Xia09, Chen09, Sakamoto09,Krupin05, Nuber08, Oguchi09, Frantzeskakis11, Ast07, Meier08,Vajna12, kmiyamoto12}.  
 The spin orientation of such low-dimensional states is locked with their crystal momenta. 
 In an ideal two-dimensional electron gas, the spins are completely oriented in-plane and orthogonal to the electron momentum~\cite{Bychkov84}.
In real systems, however, the crystal surface symmetry strongly influences the surface electronic structure and thus modifies the spin orientation~\cite{Sakamoto09, Oguchi09, Simon10, Vajna12, Frantzeskakis11, Fu09}.
   For example, Bi/Ag(111) with $C_{3v}$ symmetry exhibits an anisotropic electronic structure with a largely Rashba-spin-split band and a substantial out-of-plane spin component~\cite{Ast07, Meier08}.
  Such an anisotropic Rashba effect is driven by both in-plane and normal-to-plane crystal potential gradients as well as admixture of bulk continuum states. 
These effects have been described by {\bf\itshape k$\cdot$p} perturbation theory using an effective Rashba Hamiltonian, which takes into account the specific point group symmetry~\cite{Vajna12}. 
  Warping effects for topological insulators like Bi$_{2}$Te$_{3}$ might also be explained in the same way \cite{Fu09, Souma11}.
This shows that the crystal symmetry and the orbital character of the spin-split bands are closely related to peculiar spin structures.
 So far,  most of the Rashba systems and topological insulators are $sp$-electron materials with $C_{3v}$ point group symmetry ~\cite{Hasan10, Shitade09, Hoesch04, Hirahara08, Kadono08, Dil09, Fu09, Souma11, Kuroda10, Hsieh09, Kuroda101, Xia09, Chen09, Krupin05, Nuber08, Frantzeskakis11, Ast07, Meier08,Vajna12}. 
 
 Recently, we found nearly massless and strongly spin-polarized surface-state electrons in a spin-orbit induced symmetry gap of W(110) \cite{kmiyamoto12}. 
In contrast to the systems described above, this surface state is formed by $d$ electrons and the surface structure has $C_{2v}$ symmetry. 
  Moreover, the constant-energy cuts of this Dirac-cone-like state are found to be strongly distorted compared with $sp$-electron-based surface Dirac cones at the (111) surface plane of some Bi-based chalcogenides with $C_{3v}$ symmetry. 

\begin{figure}
\includegraphics{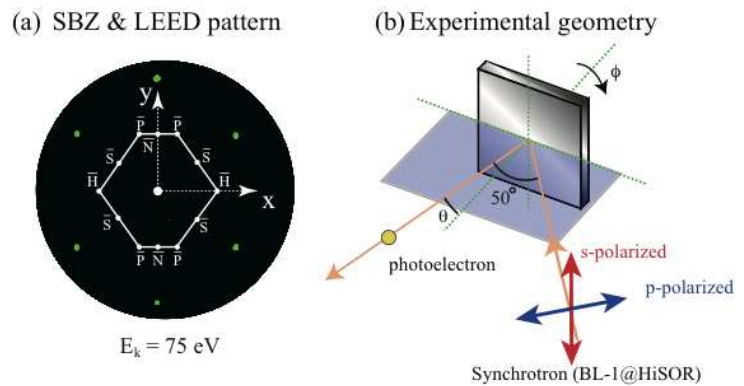}
\caption{\label{fig:epsart}(color online) (a)  LEED pattern of the clean W(110) surface and surface Brilliouin zone (SBZ) of bcc(110) with high symmetry points $\overline{\rm \Gamma}$, $\overline{\rm N}$, $\overline{\rm P}$, $\overline{\rm S}$,  and $\overline{\rm H}$.
(b) Experimental geometry for ARPES using linearly polarized synchrotron radiation. 
}
\end{figure}
\begin{figure*}
\includegraphics{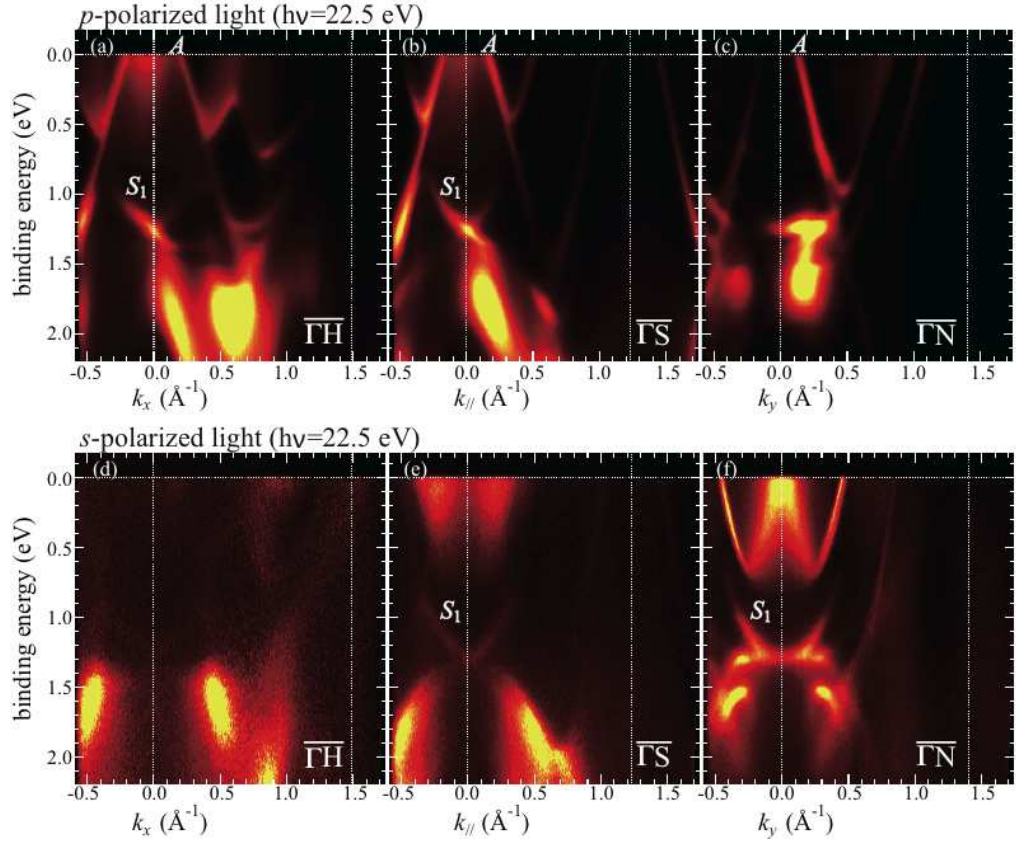}
\caption{\label{fig:epsart}(color online)  ARPES results for W(110) along $\overline{\Gamma \rm H}$, $\overline{\Gamma \rm S}$, and $\overline{\Gamma \rm N}$ excited by $p$-polarized ((a)-(c)) and $s$-polarized ((d)-(f)) light of $h\nu$ = 22.5 eV.
Horizontal dashed lines mark the Fermi level, vertical dashed lines denote high-symmetry points $\bar{\Gamma}$, $\bar{\rm H}$, $\bar{\rm S}$,  $\bar{\rm N}$, respectively. }
\end{figure*}
\begin{table}
\caption{\label{tab:example} Possible initial-state symmetries for excitation with $p$- and $s$-polarized light, according to dipole selection rules ignoring spin-orbit interaction. Note, $\Sigma_2$ type symmetry can  only be observed in an off-normal emission experiment excited by $s$-polarized light.
}
\begin{ruledtabular}
\begin{tabular}{c|cc}
 &$\overline{\Gamma \rm H}$  & $\overline{\Gamma \rm N}$ \\ \hline \\[-5pt]
$p$-pol. &$s$, $p_z$, $d_{z^2}$ : $\Sigma_{1}$ & $s$, $p_z$, $d_{z^2}$ : $\Sigma_{1}$ \\ [+1pt]

(even sym.) & $p_x$, $d_{zx}$ : $\Sigma_{3}$ 
& $p_y$, $d_{yz}$ : $\Sigma_{4}$\\[-5pt]

\\ \hline \\[-5pt]
$s$-pol. &$d_{xy}$ : $\Sigma_{2}$ (off-normal) & $d_{xy}$ : $\Sigma_{2}$ (off-normal) \\ [+1pt]

(odd sym.) & $p_y$, $d_{yz}$ : $\Sigma_{4}$ 
& $p_x$, $d_{zx}$ : $\Sigma_{3}$\\[+3pt]

\end{tabular}
\end{ruledtabular}
\end{table}
In this letter, we clarify the orbital symmetry of the spin-polarized Dirac-cone-like surface state of W(110) by polarization-dependent ARPES using $s$- and $p$-polarized synchrotron radiation. 
Furthermore, we present model calculations using an effective Rashba Hamiltonian as shown in Refs. \onlinecite{Vajna12, Simon10}.
They perfectly describe our experimentally obtained energy dispersions of the surface state along three symmetry lines $\overline{\Gamma \rm H}$, $\overline{\Gamma \rm S}$, and $\overline{\Gamma \rm N}$ and, thus, enable us to reconstruct its three-dimensional $''\rm{energy~shape}''$ as a function of the electron momentum parallel to the surface.
These findings for W(110) as a model system provide a pathway to future studies on strongly correlated $d$-electron-based topological insulators with highly anisotropic Fermi surface of $C_{2v}$ symmetry.

  A clean surface of W(110) was obtained by repeated cycles of heating in an oxygen atmosphere ($2 \times 10^{-6}$ to $2\times 10^{-7}$Pa) at 1500~K and a subsequent flash to 2300~K \cite{Bode07_preparation}. During the flash, the pressure stayed below $5 \times 10^{-7}$ Pa. 
   This cleaning procedure was effective to remove contaminants such as carbon and oxygen from the surface as confirmed by Auger electron spectroscopy as well as low-energy electron diffraction (LEED).  A very sharp (1x1) LEED pattern of the clean surface is shown in Fig. 1 (a).
Superimposed on the LEED pattern is the surface Brillouin zone (SBZ) of the bcc (110) surface. 
  The  $x$- and $y$- axes are defined parallel to the $\overline{\Gamma \rm H}$ and $\overline{\Gamma \rm N}$ symmetry lines, respectively. 
  These symmetry lines are part of mirror planes of the crystal, while $\overline{\Gamma \rm S}$ is not. 
 ARPES experiments were performed using linearly polarized undulator radiation at the beamline (BL-1) of the Hiroshima Synchrotron Radiation Center (HiSOR). 
 Electric field vectors can be switched between parallel ($p$-polarization) and perpendicular ($s$-polarization) to the plane spanned by the surface normal and photoelectron propagation vectors as schematically shown in  Fig. 1(b).  
 The angle of light incidence was $50^{\circ}$ relative to the lens axis of the electron analyzer. 
Overall experimental energy and angular resolutions were set to 15~meV and 0.1$^{\circ}$, respectively. 
 All measurements were performed at a sample temperature of 100 K.


 Figure 2 shows energy dispersion curves taken with $p$- and $s$-polarized light of $h\nu$ = 22.5 eV as a function of $k_{\parallel}$ along $\overline{\rm \Gamma H}$, $\overline{\rm \Gamma S}$ and $\overline{\rm \Gamma N}$.
Two characteristic features $A$ and $S_{1}$ were identified as surface states in former ARPES studies~\cite{Rotenberg98, kmiyamoto12}. 
Besides, broad structures are found for binding energies ($E_B$) higher than 1.35 eV.
 For excitation with $p$-polarized light, shown in Figs. 2(a)-(c), certain features are essentially independent of the $k_{\parallel}$ direction: the surface state $A$ and high-intensity broad structure with steep downward dispersion for $E_B$ $>$ 1.35 eV. 
In contrast, the Dirac-cone-like surface state $S_1$ with a crossing point at $E_{B}$ = 1.25~eV is clearly observed along $\overline{\Gamma \rm H}$ and $\overline{\Gamma \rm S}$,  whereas it cannot clearly be  resolved along $\overline{\Gamma \rm N}$.
  
   A completely different situation is observed for excitation with $s$-polarized light, shown in Figs. 2(d)-(f).
The surface state $A$ is totally absent for all the symmetry lines and a new high-intensity feature with downward dispersion appears  away from $\bar{\Gamma}$ point for $E_B$ $>$ 1.35 eV.  
The surface state $S_1$ is not observed along $\overline{\Gamma \rm H}$. 
Along $\overline{\Gamma \rm S}$, it appears with lower intensity as compared with the $p$-polarized case (Fig.~2 (b)). 
Along $\overline{\Gamma \rm N}$,  $S_1$ is clearly observed but the linear dispersion is lost around the $\bar{\Gamma}$ point. 
It is rather flat near $\bar{\Gamma}$ ($|k_{y}|$~$<$~0.2~$\rm \AA^{-1}$) and exhibits strong dispersion only for larger momenta.  
The surface state $S_1$ behaves remarkably different depending on light polarization and symmetry direction.   

We start the discussion on our ARPES results with a symmetry analysis. 
We focus on the  $\overline{\Gamma \rm H}$ and $\overline{\Gamma \rm N}$ high-symmetry directions, where mirror planes coincide with the measurement plane (see Fig. 1(b)). 
To a first approximation, i.e. when no spin-orbit coupling is considered, we can assign odd or even symmetry with respect to each mirror plane to initial states, as shown in Table I.   
In general, spin-orbit coupling promotes an intermixing between odd and even symmetry bands and thus  both kinds of states are observable in principle for both polarizations. 
Nevertheless, we observed spectral features, which appear alternately for $s$- and $p$-polarized light along $\overline{\Gamma \rm H}$ and $\overline{\Gamma \rm N}$. 
This implies that the intermixing effect is not dominant in the present case.
Therefore, we restrict our symmetry analysis to the single group symmetry ($\rm {\Sigma_1 - \Sigma_4} $).    
The wealth of experimental data in our study enables us to extend the symmetry analysis reported in Ref. \onlinecite{Gaylord87}.
      
 (i) $A$ is observed for $p$-polarized light along both $\overline{\Gamma \rm H}$ and $\overline{\Gamma \rm N}$, but not for $s$-polarized light. 
Therefore, according to Table I, we assign $\rm {\Sigma_1}$-type symmetry to $A$.

(ii) $S_1$ is observed for $p$-polarized light along $\overline{\Gamma \rm H}$ and for $s$-polarized light along  $\overline{\Gamma \rm N}$.
It is not clearly seen in the data for $s$-polarized along $\overline{\Gamma \rm H}$ and for $p$-polarized light along $\overline{\Gamma \rm N}$. 
The dispersion cannot be followed in the latter cases.
As a consequence, predominant $\rm {\Sigma_3}$-type symmetry is assigned to $S_1$ with possible minor admixture of $\Sigma_1$ ($\Sigma_2$) along $\overline{\Gamma \rm N}$ ($\overline{\Gamma \rm H}$).

(iii) The situation with the bulk continuum states for $E_{B}$ $>$ 1.35 eV is more complex. 
Certainly, spectral features appear for $p$-polarized light along all symmetry directions, while others are observed for $s$-polarized light, yet with changing intensities.
Nevertheless, from these observations, we assign predominant $\Sigma_1$- and $\Sigma_2$-type symmetries to these bulk states, while other symmetries cannot be excluded.

\begin{figure*}
\includegraphics{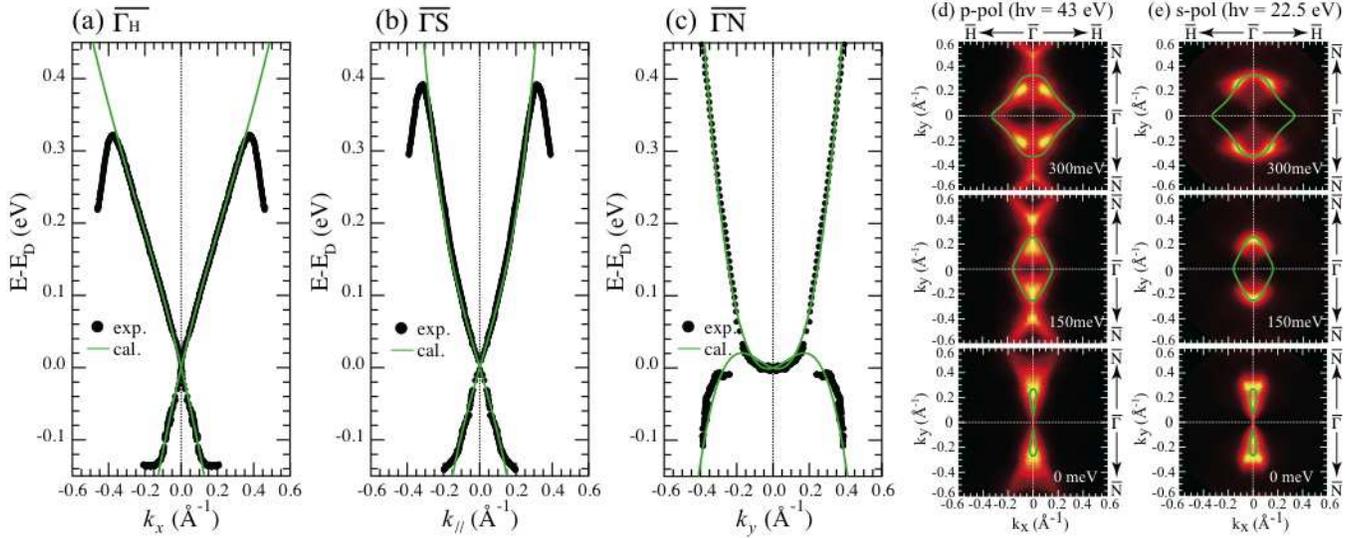}
\caption{\label{fig:epsart}(color online) (a)-(c) Intensity maxima plots for $S_{1}$ along $\overline{\Gamma \rm H}$, $\overline{\Gamma \rm S}$, $\overline{\Gamma \rm N}$ experimentally obtained (filled circles) from momentum and energy distribution curves taken at $h\nu$ = 22.5 eV (see Fig. 2). Solid lines (green) denote band dispersions obtained by effective Hamiltonian calculations considering $C_{2v}$ symmetry \cite{Vajna12}. (d)-(e) Constant energy contours at 300 meV, 150 meV, and  0 meV above the crossing point, obtained with $p$-polarized synchrotron radiation light of $h\nu$=43~eV (d) and $s$-polarized synchrotron radiation light of $h\nu$=22.5~eV (e).  The solid lines (green) are calculated results.
 }
\end{figure*}
\begin{figure}
\includegraphics{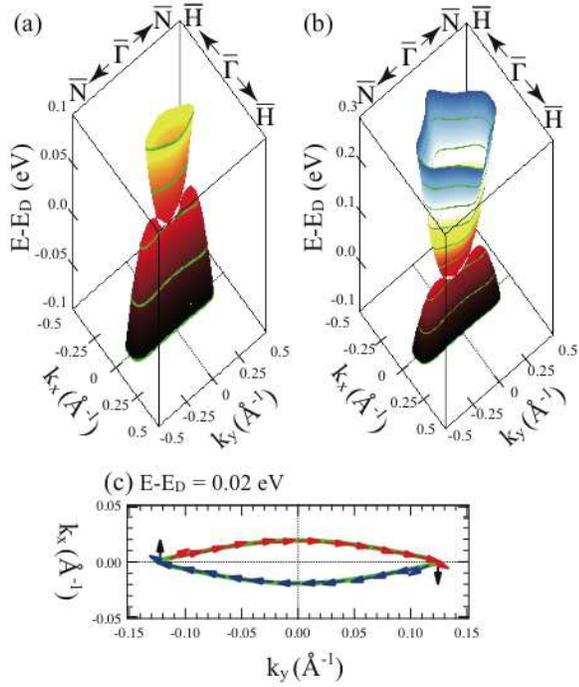}
\caption{\label{fig:epsart}(color online)  Energy contours of the surface state $S_1$ as function of $k_x$, $k_y$ for energies around the crossing point ($E_{\rm D}$) based on the parameters of the model calculation: (a) -0.1 eV $<$ $E_{\rm D}$ $<$ 0.1 eV, (b) -0.1 eV $<$ $E_{\rm D}$ $<$ 0.3 eV.
(c) spin texture for the constant energy surface (solid line) at  20 meV above $E_{\rm D}$ obtained from the model calculation. The in-plane spin components are indicated as arrows. Here, the red (blue) color means spin component parallel to positive (negative) wave vector along $\overline{\Gamma \rm N}$ .}
\end{figure}

After this symmetry analysis, we will discuss the band dispersion of the spin-polarized surface state $S_1$. 
While the dispersion is Dirac-cone-like along  $\overline{\Gamma \rm H}$ and  $\overline{\Gamma \rm S}$ with a linear behavior extending over 220 meV,  the band appears flat around $\bar{\Gamma}$ along $\overline{\Gamma \rm N}$. 
  In Figures 3 (a)-(c), we extracted the intensity maxima of $S_1$ obtained from momentum and energy distribution curves taken at $h\nu$ = 22.5 eV (Figs. 2(a), (b), and (f)).
Additional data taken at  $h\nu$ = 43 eV (not shown) yield the same dispersion curves around crossing point.
We note that this independence on photon energy confirms the two-dimensional character of the state under consideration.  

To understand the strongly anisotropic dispersion behavior of $S_1$ in more detail, we try to model our results by a simple model.
As proposed in the literature \cite{Simon10,Vajna12}, we used an effective Hamiltonian taking into account Rashba parameters up to third order and $C_{2v}$ symmetry:
 
\begin{eqnarray}
 H_{\rm eff}^{\rm {C_{2v}}}&=& \frac{\hbar^2k_x^2}{2m_{x}^*}+\frac{\hbar^2k_y^2}{2m_{y}^*}+H_{\rm R}^{(1)}+H_{\rm R}^{(3)}
\end{eqnarray}
\begin{eqnarray}
H_{\rm R}^{(1)}&=& \alpha_1^1k_x\sigma_y+ \alpha_2^1k_y\sigma_x
\end{eqnarray}
\begin{equation}
H_{\rm R}^{(3)} =  \alpha_1^3k_x^3\sigma_y+ \alpha_2^3k_x^2k_y\sigma_x+\alpha_3^3k_xk_y^2\sigma_y+ \alpha_4^3k_y^3\sigma_x
\end{equation}
 Here, $H_{\rm R}^{(1)}$ and $H_{\rm R}^{(3)}$ are the first-order and third-order perturbation Hamiltonians taken from Ref. \onlinecite{Simon10, Vajna12}.   
 The first-order Hamiltonian was sufficient to describe the free-electron-like surface state at Au(110) with parabolic dispersion \cite{Simon10}, which is almost not influenced by bulk bands.
 A third-order Hamiltonian was needed to model the bulk contributions to the surface-state behavior in Bi/Ag(111) \cite{Vajna12}.
Furthermore, the authors proposed a third-order Hamiltonian for systems with $C_{2v}$ symmetry, able to account for bulk contributions in those systems.
In the present case of W(110), the $d$-type surface state of $\Sigma_3$ symmetry with Dirac-cone-like behavior is very close in energy to bulk bands of  $\Sigma_1$ and  $\Sigma_2$ symmetries.
 Therefore, we expect that our data can be only described by first- and third-order Hamiltonians. 

 We started to fit the data along $\overline{\Gamma \rm H}$ (Fig. 3(a)) to get $m_x^*$ =  -4.7$m_0$, $\alpha_1^1$ = 1.05 eV~$\rm \AA$, and $\alpha_1^3$ =  1.13 eV~$\rm \AA^3$, where $m_0$ is free electron mass.   
We did the same for the data along  $\overline{\Gamma \rm N}$ (Fig. 3(c)) and received estimates for $m_y^*$ = 3.3$m_0$, $\alpha_2^1$ =  -0.08 eV~$\rm \AA$, and $\alpha_4^3$ = 5.57 eV~$\rm \AA^3$.
 The two remaining parameters $\alpha_2^3$ = -25 eV~$\rm \AA^3$ and $\alpha_3^3$ = 12.3 eV~$\rm \AA^3$ are determined by fitting the dispersion data along $\overline{\Gamma \rm S}$ (Fig. 3 (b)) and additional data along $\overline{\Gamma \rm P}$ (not shown).
 The fit results, included in Figs. 3 (a)-(c) as solid lines, agree well with the experimental data in the energy region around the crossing point.   
 In addition, Figs. 3(d) and (e) display constant energy contours at 300 meV, 150 meV and 0 meV above the crossing-point energy $E_{\rm D}$, obtained with $p$- and $s$-polarized light, together with calculated constant energy contours of the surface state (solid lines).
 Please note that the data for $p$-polarized light were taken at $h\nu$ = 43 eV, at which the surface-state intensity is more pronounced. 
    
 An evaluation of the obtained Rashba parameters leads to the following conclusions. 
(i) the first-order parameters exhibit a strong anisotropy ($|\alpha_1^1|$ $\approx$ 13$|\alpha_2^1|$), even larger than for Au(110) ($|\alpha_1^1|$ $\approx$ 5$|\alpha_2^1|$).  
(ii) the third-order parameters, which could be neglected for Au(110), are significant contributions to the model calculation for W(110).   
 Here, the $\Sigma_1$, $\Sigma_2$ states of the bulk continuum are located only 0.1 eV below the crossing point, which is much shallower than the case of Au(110) ($\approx$ 0.8 eV)  \cite{Simon10}.
 We assume that a spin-orbit induced mixture of $\Sigma_1$, $\Sigma_2$ states with the $\Sigma_3$ surface state leads to the strongly anisotropic spin-dependent energy splitting for different symmetry directions.

 With the model parameters derived from the experimental data, we are able to develop the energy contour of the surface state $S_1$ as a function of the $k_x$-$k_y$ plane.
Figures 4(a) and 4(b) show the strongly anisotropic shape of the contour with Dirac-cone-like behavior along $\overline{\Gamma \rm H}$ and the flat dispersion perpendicular to it ($\overline{\Gamma \rm N}$). 

  We want to emphasize that the Hamiltonian used in our model calculation includes information about the spin structure of the surface state.
 Figure 4(c) shows the spin texture (arrows) for the flattened constant energy surface (solid line) at 20 meV above $E_{\rm D}$. 
 Interestingly, the spins are almost oriented along the $\overline{\Gamma \rm N}$ direction without out-of-plane spin component.
 This situation differs from well-known Rashba systems and topological insulators with $C_{3v}$ symmetry that would approach surface states with ideal helical spin texture and shorten the spin relaxation time in the presence of disorder at the surface \cite{Souma11, Fu09}. 
 On the other hand, the spin texture along the $\overline{\Gamma \rm N}$  direction at the surface with $C_{2v}$ symmetry would generate a quasi-one dimensional edge current along the $\overline{\Gamma \rm H}$ direction.
  This is similar to the situation of a weak topological insulator, which would substantially enhance the spin relaxation time through a persistent spin helix mechanism \cite{Fu07, Bernevig06}.

 In conclusion, the orbital symmetry of the spin-polarized anisotropic Dirac-cone-like state at the W(110) surface has been examined by high-resolution ARPES with $s$- and $p$-polarized synchrotron radiation. The dominant orbital symmetry for this peculiar surface state is described as  $\Sigma_3$ ($d_{xz}$) in single group representation ignoring spin-orbit coupling. 
In our analysis, we adapted model calculations with an effective Hamiltonian including first- and third-order Rashba parameters.
 By fitting our data, we determined the parameters relevant for the model.  
The experimental surface state dispersions in three symmetry directions as well as the constant energy contours are well described by the model. 
 This suggests that the anisotropy of the surface state dispersion originates from the influence of bulk continuum states with $\Sigma_1$ and $\Sigma_2$ symmetries on the $\Sigma_3$-type surface state.  
This anisotropic $d$-band surface state of W(110) may serve as a model system of a $d$-electron based weak topological insulator constructed by stacking of one-dimensional topological edge states along a specific direction, which would enhance a spin lifetime substantially.


We thank  J. Henk and T. Shishido for stimulating discussions. M.D. gratefully acknowledges the hospitality of the Hiroshima Synchrotron Radiation Center, and support by the Japan Society for the Promotion of Science (Invitation Program for Advanced Research Institutions in Japan). The measurements were performed with the approval of the Proposal Assessing Committee of HSRC (Proposal Nos. 10-A-27, 10-B-14).

\bibliography{apssamp}

\begin{thebibliography}{60}

\bibitem{Bychkov84} Y.A. Bychkov and E. I. Rashba, JETP Lett. {\bf 39}, 78 (1984).

\bibitem{Datt90} S. Datta, and B. Das, Appl. Phys. Lett. {\bf 56}, 665 (1990).

\bibitem{Nita97} J. Nitta, T. Akazaki, H. Takayanagi, and T. Enoki, Phys. Rev. Lett.  {\bf 78}, 1335 (1997).

\bibitem{Miron10} I. M. Miron et al., Nature Mater. {\bf 9}, 230 (2010).


\bibitem{Hasan10}  M. Z. Hasan, and C. L. Kane, Rev. Mod. Phys. {\bf 82}, 3045 (2010).

\bibitem{Shitade09} A. Shitade, H. Katsura, J. Kune$\check{\rm s}$, X.-L. Qi, S.-C. Zhang, and N. Nagaosa, Phys. Rev. Lett {\bf 102}, 256403 (2009).



\bibitem{Hoesch04} M. Hoesch et al., Phys. Rev. B {\bf 69}, 241401(R) (2004).

\bibitem{Hirahara08} T. Hirahara et al., New J. Phys. {\bf 10}, 083038 (2008).

\bibitem{Kadono08} T. Kadono et al., Appl. Phys. Lett. {\bf 93}, 252107 (2008).


\bibitem{Dil09} J. H. Dil, J. Phys. Condens. Matter {\bf 21}, 403001 (2009).

\bibitem{Krupin05} O. Krupin et al., Phys. Rev. B {\bf 71}, 201403(R) (2005).

\bibitem{Nuber08} A. Nuber, M. Higashiguchi, F. Forster, P. Blaha, K. Shimada, and F. Reinert, Phys. Rev. B {\bf 78}, 195412 (2008).


\bibitem{Sakamoto09} K. Sakamoto, et al., Phys. Rev. Lett {\bf 102}, 096805 (2009).


\bibitem{Frantzeskakis11} E. Frantzeskakis, and M. Grioni, Phys. Rev. B {\bf 84}, 155453 (2011).

\bibitem{Ast07} C. R. Ast et al., Phys. Rev. Lett {\bf 98}, 186807 (2007).

\bibitem{Meier08} F. Meier, H. Dil, J. Lobo-Checa, L. Patthey, J. Osterwalder, Phys. Rev. B {\bf 77}, 165431 (2008).


\bibitem{Souma11} S. Souma et al., Phys. Rev. Lett {\bf 106}, 216803 (2011).


\bibitem{Hochstrasser02}M.~Hochstrasser, J. G.~Tobin, E.~Rotenberg, and S. D.~Kevan, Phys. Rev. Lett. {\bf 89}, 216802 (2002).


\bibitem{Xia09} Y. Xia et al., Nature Phys. {\bf 5}, 398 (2009).





\bibitem{Chen09} Y. L. Chen et al., Science {\bf 325}, 178 (2009).

\bibitem{Kuroda10}K. Kuroda et al., Phys. Rev. Lett. {\bf 105}, 146801 (2010). 

\bibitem{Hsieh09}  D. Hsieh et al., Nature {\bf 460}, 1101 (2009).

\bibitem{Kuroda101} K. Kuroda et al., Phys. Rev Lett {\bf 105}, 076802 (2010).

\bibitem{kmiyamoto12}K. Miyamoto et al., Phys. Rev. Lett. {\bf 108}, 066808 (2012). 

\bibitem{Simon10}E. Simon, A. Szilva, B. Ujfalussy, B. Lazarovits, G. Zarand, and L. Szunyogh, Phys. Rev. B {\bf 81}, 235438 (2010).  

\bibitem{Vajna12}Sz. Vajna, E. Simon, A. Szilva, K. Palotas, B. Ujfalussy, and L. Szunyogh, Phys. Rev. B {\bf 85}, 075404 (2012). 

\bibitem{Fu09} L. Fu, Phys. Rev. Lett. {\bf 103}, 266801 (2009).

\bibitem{Oguchi09} T. Oguchi, and T. Shishidou, J. Phys.: Condens. Matter {\bf 21}, 092001 (2009).




\bibitem{Rotenberg98} E.~Rotenberg and S. D.~Kevan, Phys. Rev. Lett.  {\bf 80}, 2905 (1998).

\bibitem{Gaylord87} R. H. Gaylord and S. D. Kevan, , Phys. Rev. B {\bf 36}, 9337 (1987).






\bibitem{Shikin08} A. M. Shikin et al., Phys. Rev. Lett. {\bf 100}, 057601 (2008).

\bibitem{Rotenberg08}E.~Rotenberg, O.~Krupin, and S. D.~Kevan, New J. Phys. {\bf 10}, 023003 (2008).


\bibitem{Bode07_preparation} M. Bode, S. Krause, L. Berbil-Bautista, S. Heinze, and R. Wiesendanger, Sur. Sci. {\bf 601}, 3308 (2007).

\bibitem{Fu07} L.Fu, C. L. Kane, and E. J. Mele,  Phys. Rev. Lett. {\bf 98}, 106803 (2007). 

\bibitem{Bernevig06}B. A. Bernevig, J. Orenstein, and S-C. Zhang, Phys. Rev. Lett. {\bf 97}, 236601 (2006).














%
\end{thebibliography}

\end{document}